# Non-radiative decay and stability of *N*-heterocyclic carbene iridium(III) complexes

Xiuwen Zhou, Benjamin J. Powell

*School of Mathematics and Physics, The University of Queensland, Brisbane, Queensland 4072, Australia*

**ABSTRACT**

Devices based on deep-blue emitting iridium (III) complexes with *N*-heterocyclic carbene (NHC) ligands have recently been shown to give excellent performance as phosphorescent organic light-emitting diodes (PHOLEDs). To facilitate the design of even better deep-blue phosphorescent emitters we carried out density functional theory (DFT) calculations of the lowest triplet ($T_1$) potential-energy surfaces (PES) upon lengthening the iridium-ligand (Ir-C) bonds. Relativistic time dependent-DFT (TDDFT) calculations demonstrate that this changes the nature of $T_1$ from a highly-emissive metal-to-ligand charge transfer ($^3$MLCT) state to a metal centered ($^3$MC) state where the radiative decay rate is orders of magnitude slower than that of the $^3$MLCT state. We identify the elongation of an Ir-C bond on the NHC group as the pathway with lowest energy barrier between the $^3$MLCT and $^3$MC states for all complexes studied and show that the barrier height is correlated with the experimentally measured non-radiative decay rate. This suggests that the thermal population of $^3$MC states is the dominant non-radiative decay mechanism at room temperature. We show that the $^3$MLCT -> $^3$MC transition is reversible, in marked contrast to other deep blue phosphors containing coordinating nitrogen atoms, where the population of $^3$MC states breaks Ir-N bonds. This suggests that, as well as improved efficiency, blue PHOLEDs containing phosphors where the metal is only coordinated by carbon atoms will have improved device lifetimes.

## INTRODUCTION

A major challenge in the field of phosphorescent organic light-emitting diode (PHOLED) technology is the lack of highly efficient emitters for saturated blue and deep-blue color to meet the demand of high quality displays.[1] Extensive efforts to develop blue phosphorescent iridium(III) [Ir(III)] complexes have been motivated by the existing highly efficient green and red Ir(III) phosphors, especially the green emitter *fac*-tris(2-phenyl-pyridyl)iridium(III) [Ir(ppy)$_3$],[2] which displays a near unity photoluminescence quantum yield (PLQY). However, there are still only a small number of known blue phosphorescent Ir(III) complexes. Furthermore, most of these emit sky-blue light, which is inadequate for full color displays an lighting applications. On the other hand, the known deep blue emitters suffer from relatively low PLQY and stability issues in

devices.[1]

Ligand modification of existing emitters is often used to tune the color of Ir(III) complexes, however it also often dramatically changes the room-temperature PLQY; many times from near 1 to near 0.[3-7] The dramatic decrease in PLQY is primarily[3, 6] due to the increase in non-radiative rate of the emissive state (typically from ~$10^3$ s$^{-1}$ to >$10^8$ s$^{-1}$), with the radiative rate remaining of the order of $10^5$ s$^{-1}$ for the most promising deep blue emitters.[3, 6] Therefore, in order to predict the PLQY of Ir(III) complex, the key theoretical challenge is to understand the non-radiative decay processes of the emissive state. A number of non-radiative processes have been discussed including: intramolecular vibronic coupling quenching[8-13], thermal population to non-radiative states[6, 14-26], intermolecular quenching[27-30].

The quantum efficiency can be expressed as a function of radiative and non-radiative rates by[31],

$$\Phi_{PL} = \frac{k_r}{k_r + k_{nr}}, \qquad (1)$$

where $\Phi_{PL}$ is the PLQY, $k_r$ is the radiative rate, $k_{nr}$ is the non-radiative rate. Different non-radiative decay mechanisms lead to dramatically different temperature dependences of $k_{nr}$. For example, $k_{nr}$ due to vibronic coupling is largely temperature independent, whereas the rate due to the thermal population to an upper level non-radiative state (which cannot decay radiatively or does so only very slowly) is highly temperature dependent. This non-radiative process would lead to an activated behavior, hence the rate can be expressed in terms of the Arrhenius equation[19],

$$k_{nr}(T) = k_a e^{-\Delta E/k_B T}, \qquad (2)$$

where $k_a$ is the rate constant, $\Delta E$ is the activation energy to the upper level, $k_B$ is the Boltzmann constant and the ellipses represent other possible contributions. The exponential decrease in $k_{nr}(T)$ as $\Delta E$ increases implies that small differences in the activation energies change $k_{nr}(T)$ dramatically. Several authors[6, 14, 15, 17-21, 23, 25, 26] have argued that the thermal population of non-radiative metal-centered ($^3$MC or $^3$dd$^*$) states is the dominant non-radiative decay mechanism in blue phosphorescent Ir(III) complexes at room temperature. Since these excitations have strong metal-to-metal character (d -> d$^*$ transition) and they provide little possibility of radiative decay. This is consistent with the experimental observation that $k_{nr}$, and hence the PLQY, is clearly temperature dependent, i.e., the low-temperature $k_{nr}$ is very low (PLQY is very high) for many complexes while on increasing the temperature towards room-temperature $k_{nr}$ (PLQY) increases (decreases) dramatically[6]. For example, we have shown[25] that a there are only small energy barriers (relative to $k_BT$) between the highly emissive metal to ligand charge transfer ($^3$MLCT) states and non-radiative $^3$MC states in a family of Ir(III) complexes based on *fac*-tris(1-methyl-5-phenyl-3-*n*-propyl-[1,2,4]triazolyl)iridium(III) [Ir(ptz)$_3$]. The lowest energy barrier path to non-radiative states for these complexes is the elongation of an Ir-N bond. The calculated barrier energy correlated with the experimental non-radiative rate, as expected from Eq. (2). Moreover, once the non-radiative states were formed, the complexes were not able to return to the emissive states, resulting in permanent damage to the structure via breaking of the Ir-N bond, which explains the inferior stability of PHOLED devices based on them. These findings and the applied theoretical methods have been used to predict Ir(III) complexes with higher PLQY in the design of new complexes based on Ir(ptz)$_3$.[26]

Recently, deep-blue-emitting PHOLEDs based on tris-(*N*-phenyl,*N*-methyl-pyridoimidazol-2-yl)iridium(III), Ir(pmp)$_3$ (see right panels of Figure 1) achieved the brightest deep blue emission among the PHOLEDs reported. Indeed, devices based on its facial isomer [*fac*-Ir(pmp)$_3$] perform close to the most stringent US National Television System Committee requirements.[32] The meridional isomer (*mer*-Ir(pmp)$_3$) was also found equally efficiently luminescent; the PLQYs of the two isomers are equal within the reported errors, cf. Table 3. However, both isomers of the closely related tris-(*N*-phenyl, *N*-methyl-benzimidazol-2-yl) Ir(III), Ir(pmb)$_3$ (see left panels in Figure 1) have much lower luminescence efficiency at room temperature, in fact *mer*-Ir(pmb)$_3$ is barely luminescent[33] (the experimental PLQYs are reproduced in Table 3). In order to further design new blue Ir(III) complexes based on NHC ligands suitable as emitters in PHOLEDs, it is very important to understand the non-radiative pathways and stability of these complexes, and why apparently small chemical changes so dramatically alter their photophysical properties.

As a first step towards these goals, here we investigate the possible reaction paths to non-radiative states for these four complexes and the correlation between the lowest activation barrier to non-radiative states and the experimental non-radiative rate. We also discuss how ligand substitution and structural isomerization affect the photoluminescence efficiency and the stability of these complexes. We show that the barriers between the $^3$MLCT and $^3$MC excited states are large for both isomers of the Ir(pmp)$_3$ complex, consistent with their high PLQYs, and smaller in Ir(pmp)$_3$. Furthermore we find that, unlike in other blue phosphorescent complexes, when the material reaches the $^3$MC state this does *not* break the stretched metal-ligand bond. This is allows for the possibility of complex returning to the emissive $^3$MLCT state, further increasing the PLQY of these complexes. It also means that excitations of the $^3$MC state do not cause permanent damage to the complex – this suggests that PHOLEDs based on Ir(pmp)$_3$ should have much better device lifetimes than those for PHOLEDs with other blue phosphorescent active materials.

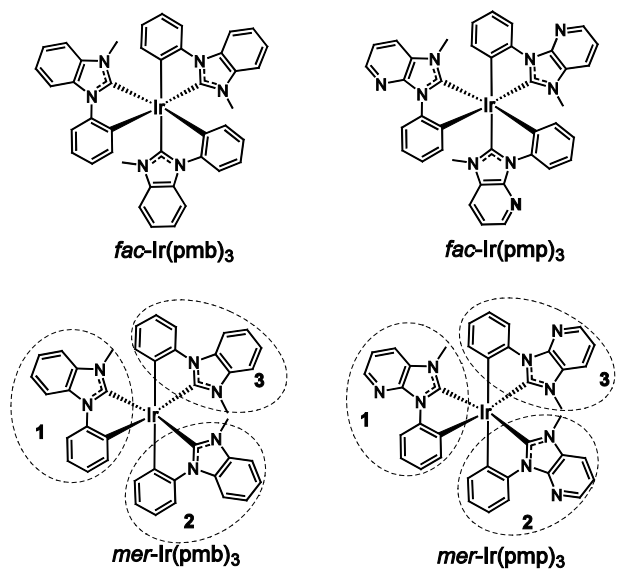

Figure 1. Structures of the facial (*fac*-) and meridional (*mer*-) isomers of Ir(pmb)$_3$ and Ir(pmp)$_3$. These are closely related complexes – a methane replaced by a nitrogen atom in the NHC ligands (CH -> N) from pmb to pmp. This modification not only redshifts the emission from near ultraviolet to the deep blue region, but also greatly increases the PLQY (see Table 3). The ligands of *mer*-isomers are numbered to distinguish the six inequivalent Ir-ligand bonds.

## CALCULATIONS

### Geometry

The structures of *fac-* and *mer-*Ir(pmb)$_3$ were optimized taking the measured crystal structures[33] as starting points. The structures of *fac-* and *mer-*Ir(pmp)$_3$ were then optimized with the initial structures based on the optimized structures of *fac-* and *mer-*Ir(pmb)$_3$, respectively. The geometry optimization was performed with density functional theory (DFT)[34, 35] using the B3LYP[36-38] functional. The 6-31G(d,p) basis set[39] was used for hydrogen, carbon, nitrogen, and fluorine, and the LANL2DZ basis set[40] with an effective core was used for iridium. All calculations in this work except those of the radiative rate and electronic excitations were performed using the Gaussian 09 set of programs[41], with the functional and basis set described above used throughout.

### Triplet potential-energy surface

Several previous studies[14, 15, 6, 25] indicate that for emissive cyclometalated complexes the reaction path to non-radiative states is the elongation or even breaking of a metal-ligand bond. To find the lowest-energy-barrier reaction path to non-radiative states, we therefore performed a relaxed potential-energy surface (PES) scan for the emissive state, that is, the lowest energy triplet (T$_1$) state, along the reaction path of the elongation of each inequivalent Ir-C bond. Note that, for the two *fac-*isomers (upper panels of Figure 1), each complex has two types of inequivalent Ir-ligand bonds, i.e., three equivalent Ir-C (NHC) bonds and three equivalent Ir-C (PHE) bonds, where PHE denotes phenyl and the label in brackets indicates the group containing the relevant carbon atom. However, for the two *mer-*isomers (lower panels of Figure 1), all six Ir-ligand bonds are unique. To distinguish these six bonds, we label the three ligands 1-3 (Figure 1).

In the relaxed PES scan, the starting structure had the Ir-C bond length equal to that found in the optimized ground-state structure. This bond was then lengthened in 0.1 Å increments. The geometry at each point along this path was optimized subject to the length constraint on this single extended bond. The initial geometry for each optimization after increasing the Ir-C bond length took the position of the remaining atoms to be those found in optimized structure for the previous Ir-C bond length. The energy of the T$_1$ state was then calculated. Unrestricted DFT was used for treating the T$_1$ states since they contain unpaired electrons. Unrestricted DFT is known to be quite good at modelling the properties of open-shell molecular systems including spin polarization (in contrast, the unrestricted Hartree-Fock approximation overestimates polarization) and energetics.[42] To optimize the triplet states and calculate the triplet energies, a quadratically convergent self-consistent field (SCF) procedure[43] was used that is slower than regular SCF but more reliable.

### Radiative rate and low-energy excitations

Relativistic calculations were performed for selected structures to obtain the radiative rate and more accurate information on electron excitations. In these calculations, spin-orbit coupling (SOC) was added perturbatively to one-component TDDFT[44] utilizing the one-component zeroth order regular approximation (ZORA).[45-47] A total of 40 spin-mixed excitations were calculated. The calculations were carried out with ADF (2017 version),[48] using the B3LYP[36-38] functional and Slater type TZP basis sets[49, 50] with a frozen core approximation for the iridium [1s 2s 2p 3s 3p 3d 4s 4p 4d 4f], nitrogen [1s] and carbon [1s] shells.

## RESULTS & DISCUSSION

### Optimized geometries and the concomitant electronic structure

To understand the impact of isomerization and CH -> N substitution on the ligand on the ground-state geometry, key geometric parameters (length of six Ir-ligand bonds) are collated in Table 1. The ligand substitution causes negligible change in the length of the six Ir-ligand bonds; all bond lengths are almost identical for Ir(pmb)$_3$ and Ir(pmp)$_3$. *Fac*-isomers have equal lengths for all Ir-C (NHC) bonds and all Ir-C (PHE) bonds due to their C$_3$ symmetry. The Ir-C (NHC) bonds are 0.05 Å shorter than the Ir-C (PHE) bonds. However, *mer*-isomers have different lengths for all six Ir-ligand bonds due to its asymmetric molecular structure (C$_1$). Our interest is to find the weakest Ir-C bond. We will see below, that for both facial isomers the shorter NHC bonds provide a lower energy barrier pathway from the $^3$MLCT state to the $^3$MC state. This somewhat surprising as one might naïvely expect the PHE bonds to be weaker given their greater length in the ground state geometry. This can be understood in terms of the electronic changes in the excited states – which emphasizes that the excited states behave quite differently from the ground state. Nevertheless, we will see below that in both meridional complexes the elongation of the longest Ir-C (NHC) bond in the ground state geometry, i.e., Ir-C (NHC$_3$), has the lowest barrier between the $^3$MLCT and $^3$MC states.

Table 1. Bond lengths (six Ir-C bonds) of *fac*- and *mer*- isomers of Ir(pmb)$_3$ and Ir(pmp)$_3$ at geometries optimized for ground state. *Fac*-isomers have two inequivalent types of metal-ligand bonds, i.e., Ir-C (NHC) bonds and Ir-C (PHE), while *mer*-isomers have six inequivalent bonds, the indices 1, 2, and 3 denote the ligand containing that bond (cf. Figure 1).

| complex | Ir-C (NHC$_1$) | Ir-C (PHE$_1$) | Ir-C (NHC$_2$) | Ir-C (PHE$_2$) | Ir-C (NHC$_3$) | Ir-C (PHE$_3$) |
|---|---|---|---|---|---|---|
| *fac*-Ir(pmb)$_3$ | 2.06 | 2.11 | 2.06 | 2.11 | 2.06 | 2.11 |
| *fac*-Ir(pmp)$_3$ | 2.06 | 2.11 | 2.06 | 2.11 | 2.06 | 2.11 |
| *mer*-Ir(pmb)$_3$ | 2.06 | 2.09 | 2.05 | 2.12 | 2.08 | 2.14 |
| *mer*-Ir(pmp)$_3$ | 2.06 | 2.10 | 2.05 | 2.12 | 2.08 | 2.14 |

It has been previously shown that the lowest excitation of cyclometalated phosphors is predominately a highest occupied molecular orbital (HOMO) to lowest unoccupied molecular orbital (LUMO) transition and their T$_1$ states have strong MLCT character.[51-53, 25] The NHC Ir(III) complexes studied here also have a lowest excitation that is predominantly a HOMO->LUMO transition (~ 60% weight). From the frontier molecular orbitals shown in Figure 2, the HOMO of *fac*-isomer is distributed mainly on the metal atom and the three phenyl rings, whereas its LUMO is distributed mainly on the NHC ligands. For the *mer*-isomer, the HOMO is mainly distributed on the metal and two of three phenyl rings (ligands 1 and 2; cf. Figure 1), while the LUMO is distributed mainly on the NHC part of ligand 3. Therefore, in their optimized ground state geometries, the T$_1$ states of both isomers have both metal-to-ligand change transfer character ($^3$MLCT) and ligand-to-ligand change transfer character. This charge redistribution is crucial for understand the bond strengths in the excited states.

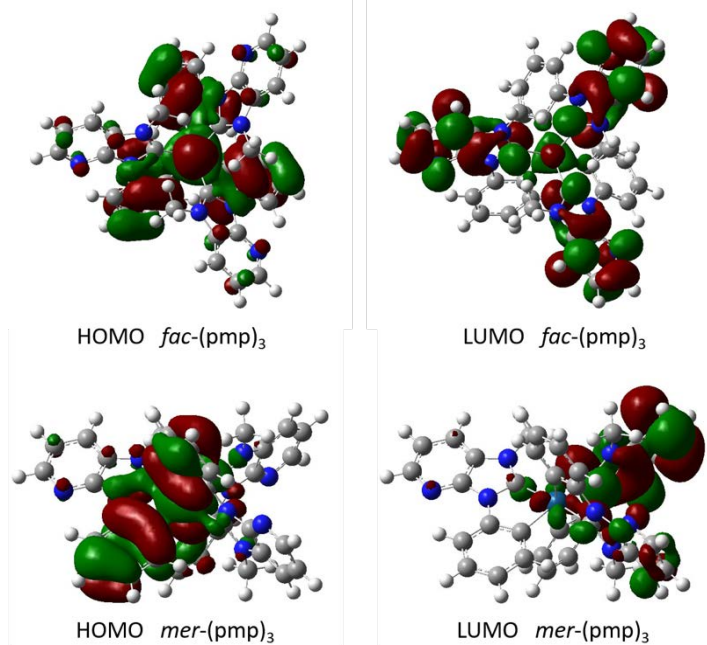

Figure 2. Frontier molecular orbitals (HOMO and LUMO) of *fac-* and *mer*-Ir(pmp)$_3$ in their optimized ground state geometries. The equivalent orbitals for *fac-* and *mer*-Ir(pmb)$_3$ show the same trend, i.e., a significant metallic contribution to the HOMO but not the LUMO.

**Possible non-radiative decay paths and characterization of $^3$MC states**

The calculated PESs for the T$_1$ state under the elongation of an Ir-ligand bond are shown in Figure 3. In order to find the non-radiative paths with the lowest energy barriers, we analyzed the character of the T$_1$ states along the reaction paths to see if any non-radiative states occurred. This allowed us to identify the lowest barrier non-radiative paths. Prototypical non-radiative states on the PES with the lowest-energy-barrier area marked in Figure 3. These states are chosen as states that have strong 3MC character, but as short an Ir-C bond as possible given this constraint. The precise choice of this state is a somewhat subjective choice. However, results for other choices are similar. The calculated results for radiative rate, excitation energy and oscillator strengths of the lowest excitations are collated in Table 2. Note that the radiative rates of the $^3$MC states (~10$^2$ s$^{-1}$) are 2-3 orders of magnitudes slower than that of the respective emissive ($^3$MLCT) states (~10$^5$ s$^{-1}$). All $^3$MC states have very low excitation energies (< 1 eV) and small oscillator strengths (~10$^5$).

For the both *fac*-isomers, the lowest barrier non-radiative path is the elongation of an Ir-C (NHC) bond rather than an Ir-C (PHE) bond. This indicates that *in the excited state* the Ir-C (NHC) bonds are weaker than the the Ir-C (PHE) bonds. This can be understood as a consequence of the charge transferred into the NHC group in the excited state weakening the Ir-C (NHC) bond. For both *mer*-isomers, the lowest-energy-barrier non-radiative path is the elongation of the Ir-C (NHC) bond in ligand 3 (cf. Figure 1), which is the longest, and thus presumably the weakest, Ir-C (NHC) bond in the ground state geometry. Interestingly, the LUMO of *mer*-isomers are also mainly distributed on the NHC part of ligand 3. This is consistent with the idea that the electron transfer from the metal and phenyl rings to this NHC part facilitates the elongation of this bond.

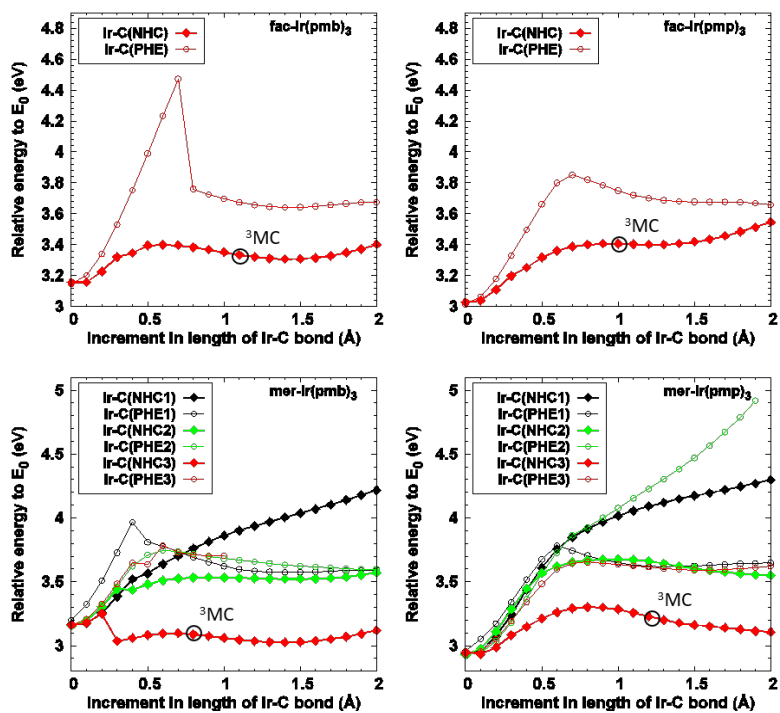

Figure 3. PESs of *fac*- and *mer*-Ir(pmp)$_3$ and *fac*- and *mer*-Ir(pmp)$_3$ as an Ir-C bond is lengthened and the remaining atomic positions are optimized self-consistently. All curves show the energy of the T$_1$ state when the geometry is optimized for the T$_1$ state. See Figure 1 for the definition of the ligand numbering. The circled data point marked "$^3$MC" denotes the prototypical non-radiative state discussed below (see, particularly, Table 2).

Table 2. Calculated average excitation energies ($\varepsilon$), oscillator strengths ($f$), and radiative rates ($k_r$) of the lowest three electronic excitations (i.e., the three substates of T$_1$) for the $^3$MLCT states (i.e., the T$_1$ states in the ground-state geometries) and the prototypical $^3$MC (non-radiative) states (marked in Figure 2) of the Ir(pmb)$_3$ and Ir(pmp)$_3$ complexes.

| complex | State | $\varepsilon$ (eV) | f | $k_r$ (s$^{-1}$) |
|---|---|---|---|---|
| *fac*-Ir(pmb)$_3$ | $^3$MLCT | 3.31 | 2×10$^{-4}$ | 9×10$^4$ |
| | $^3$MC | 0.11 | 3×10$^{-5}$ | 2 |
| *fac*-Ir(pmp)$_3$ | $^3$MLCT | 2.98 | 3×10$^{-4}$ | 1×10$^5$ |
| | $^3$MC | 0.37 | 2×10$^{-5}$ | 1×10$^2$ |
| *mer*-Ir(pmb)$_3$ | $^3$MLCT | 3.23 | 2×10$^{-4}$ | 8×10$^4$ |
| | $^3$MC | 0.97 | 1×10$^{-5}$ | 5×10$^2$ |
| *mer*-Ir(pmp)$_3$ | $^3$MLCT | 2.84 | 1×10$^{-4}$ | 5×10$^4$ |
| | $^3$MC | 0.15 | 1×10$^{-5}$ | 9 |

## Impact of CH->N ligand substitution and structural isomerization on PLQY

The lowest energy barriers to non-radiative states for all four complexes are tabulated in Table 3. A simple relationship is found between the activation barrier and the non-radiative rate of the four complexes: the larger the barrier, the lower the non-radiative rate. Clearly this is consistent with decay via a $^3$MC state being the dominant non-radiative decay pathway at room temperature for these NHC Ir(III) complexes (see Equation 1 and 2).

To understand the impact of ligand substitution (CH->N) on the luminescence efficiency, let us take a closer look at the energy barriers to non-radiative rate and the shape of PES from emissive states to non-radiative states. From *fac*-Ir(pmb)$_3$ to *fac*-Ir(pmp)$_3$, the energy barrier to non-radiative rate increases from 0.24 eV to 0.38 eV, this causes the non-radiative decay rate to decrease by an order of magnitude. However, the radiative rates (both calculated and measured) of two complexes are in the same order of magnitude. Therefore the overall PLQY is increased by the ligand substitution (CH->N). Moreover, the energy barrier from non-radiative state back to emissive state is lowered, Figure 4, which indicates that the ligand substitution leads to less stable non-radiative states.

A similar but even higher impact of ligand substitution can be observed from *mer*-Ir(pmb)$_3$ to *mer*-Ir(pmp)$_3$ -- the energy barrier to non-radiative rate increases from 0.09 eV to 0.35 eV, and the non-radiative state of *mer*-Ir(pmb)$_3$ is more stable than its emissive state (lower T$_1$ energy in the non-radiative state)! This explains why *mer*-Ir(pmb)$_3$ is nearly non-emissive at room temperature (with a very low PLQY of 0.2). The impact of CH->N substitution on k$_{nr}$ might be due to the greater electronegativity of the N atom than the CH group strengthening the Ir-C (NHC) bond and thus lifting the energy barrier to non-radiative states and resulting a slower k$_{nr}$.

The impact of isomerization (from facial to meridional) on PLQY is however less significant than that of CH->N substitution. Overall the *fac*-isomers have higher energy barriers to the non-radiative rate states and thus smaller k$_{nr}$, although the difference between *fac*-Ir(pmp)$_3$, and *mer*-Ir(pmp)$_3$ is in fact very small (the former is 0.03 eV greater in energy barrier). The lowered symmetry (from C$_3$ to C$_1$) allows the weakening of one of the Ir-C (NHC) bonds and thus decreases the energy barrier to non-radiative rate and increases the k$_{nr}$. As the weakest bond predominately determines the non-radiative decay rate any strengthening of the other bonds does not compensate for this.

Table 3. Calculated activation energy to the $^3$MC states *(ΔE)*, non-radiative rate at 295 K divided by the rate constant (k$_{nr}$[295K]/k$_a$, see Equation 2), radiative rate at 0 K (k$_r$[0K]), and emission energy (λ$_{max}$) of four NHC Ir(III) complexes. The measured room-temperature non-radiative rates (k$_{nr}$), radiative rates (k$_r$), photoluminescent quantum yields (Φ$_{PL}$), and emission peak wavelengths (λ$_{max}$) are also given for comparison.

| | calculated | | | | experimental | | | | |
|---|---|---|---|---|---|---|---|---|---|
| complex | ΔE (eV) | k$_{nr}$[295K]/k$_a$ | k$_r$[0K] (s$^{-1}$) | λ$_{max}$ (nm)* | k$_{nr}$[295K] (s$^{-1}$) | k$_r$[295K] (s$^{-1}$) | Φ$_{PL}$ [295K] | λ$_{max}$ (nm) | Ref. |
| *fac*-Ir(pmb)$_3$ | 0.24 | 6.8×10$^{-5}$ | 0.9×10$^5$ | 375 | 5.7×10$^5$ | 3.4×10$^5$ | 37 | - | 6 |
| | | | | | 4.0×10$^6$ | 1.8×10$^5$ | 4 | 380 | 33 |
| *fac*-Ir(pmp)$_3$ | 0.38 | 3.6×10$^{-7}$ | 1.0×10$^5$ | 416 | (2.0±0.4)×10$^5$ | (6.4±1.3)×10$^5$ | 76±5 | 418 | 32 |
| *mer*-Ir(pmb)$_3$ | 0.09 | 3.1×10$^{-2}$ | 0.8×10$^5$ | 384 | 6.50×10$^7$ | 1.3×10$^5$ | 0.2 | 385 | 33 |
| *mer*-Ir(pmp)$_3$ | 0.35 | 9.7×10$^{-7}$ | 0.5×10$^5$ | 437 | (2.7±0.4)×10$^5$ | (10±2)×10$^5$ | 78±5 | 465 | 32 |

*Emission wavelengths are converted from the excitation energy values of $^3$MLCT states in ground state geometry in Table 2.

**Stability of NHC Ir(III) complexes**

To understand the stability of the molecules in the $^3$MC state we investigated how the energetics of $T_1$ state and molecular structures change when the Ir-C (NHC) bond is shorted starting from the $^3$MC state marked in Figure 2 for the lowest-energy-barrier non-radiative path identified above (Figure 4).

We find that the $^3$MC states of all four complexes are able to return to a geometry very similar to the equilibrium one, which have similar $T_1$ energy (see the initial points of the PESs in Figure 4) and close metal-ligand bond lengths. This indicates that even these complexes are thermally populated to $^3$MC states, they are still able to return to the $^3$MLCT states. The Ir-C bond does not break! This is very different behavior from that found for complexes with Ir-N metal-ligand bonds.[25, 26] The thermal population to $^3$MC states does not lead to a permanent damage to the structure of the complex, although it results in an increased non-radiative rate. In contrast, Ir(ptz)$_3$ and similar complexes are unable to return to the $^3$MLCT state once the $^3$MC state is formed. Instead shortening the relevant bond length results in another $^3$MC state (one of other Ir-N bonds grows to more than 1 Å longer than the equilibrium $^3$MLCT state), which has lower $T_1$ energy than the equilibrium $^3$MLCT state! This indicates that the formation of $^3$MC state in those Ir(ptz)$_3$ based complexes leads to both an increased non-radiative rate and a permanent damage of the emitters in device. Figure 4 leads us to predict that all four of these complexes should have good stability in devices under repeated use. Thus we expect that PHOLEDs with Ir(pmp)$_3$ as their active material should have good device lifetimes.

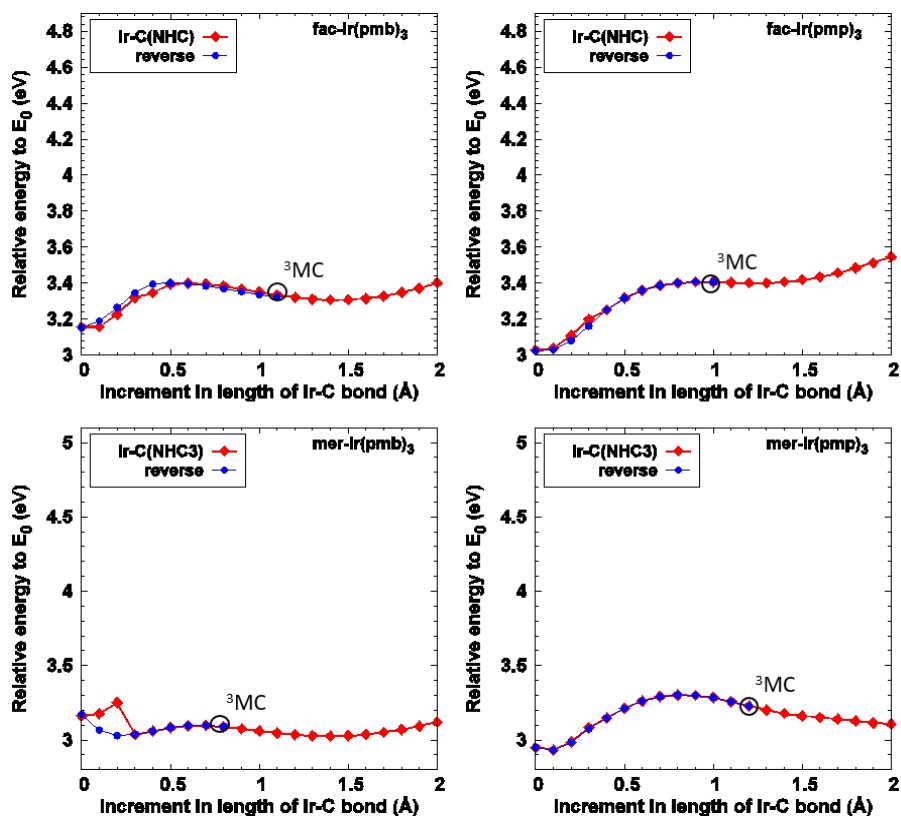

Figure 4. PESs of *fac*- and *mer*-Ir(pmp)$_3$ and *fac*- and *mer*-Ir(pmp)$_3$ as an Ir-C (NHC) bond is lengthened and the remaining atomic positions are optimized self-consistently. All curves show the energy of the T$_1$ state when the geometry is optimized for the T$_1$ state. The red diamonds denote a reaction path where an Ir-C (NHC) bond is elongated (i.e., moving from left to right) starting from the $^3$MLCT state along the lowest barrier path to $^3$MC state (cf. Figure 3), while the blue diamonds denote the reaction path on the shortening of a Ir-C (NHC) bond (i.e., moving from right to left) starting from the prototypical $^3$MC state.

## CONCLUSIONS

We have demonstrated that the experimentally measured non-radiative decay rates in a family of *N*-heterocyclic carbene iridium(III) complexes are correlated with the energy barriers between $^3$MLCT and $^3$MC excited states. Furthermore, the calculated radiative decay rates of the latter are orders of magnitude slower than the radiative decay rates of the former, which are consistent with the experimentally measured rates. This suggests that thermal population of the "dark" $^3$MC states is the dominant non-radiative decay mechanism. The T$_1$ PESs indicate that the elongation of an Ir-C (NHC) bond is the most likely path to the dark states for all four NHC Ir(III) complexes discussed in this paper. It is interesting to contrast this with ground state structure, where the Ir-C (NHC) bonds are longer than the Ir-C (PHE) bonds. However, the calculated frontier molecular orbitals suggest that there is significant charge transferred into the NHC moieties in the excited state. This appears to lead to the weakening of the Ir-C (NHC) bonds in the excited state.

CH->N ligand substitution is found to stabilize the Ir-C (NHC) bond in the excited state; lifting the activation barrier and hence suppressing the non-radiative rate. Isomerization from facial to meridional isomer lowers the symmetry of the molecule, which weakens at least one Ir-C (NHC) bond and thus increases the non-radiative rate.

Finally, we found that the lengthening of the Ir-C (NHC) bond is reversible. This is in marked contrast to Ir(ptz)$_3$ and its fluorinated derivatives, which we studied previously[25, 26]. In these materials lengthening the Ir-N bond sufficiently to reach the $^3$MC state is irreversible as it breaks the Ir-N bond. This indicates that the NHC complexes studied here are able to return to the $^3$MLCT state from the $^3$MC state. More importantly, the thermal population of complexes in the $^3$MC state does not lead to permanent damage of the complexes in a device setting.

More generally it is interesting to speculate that the stability of the Ir-C bonds in $^3$MC excited states may be significantly better than Ir-N bonds. It seems likely that, for example, the stability and efficiency of Ir(ppy)$_3$ could arise from a large barrier between the $^3$MLCT and $^3$MC states, rather than the properties of the $^3$MC state itself. This suggests that exploring other complexes where all coordinating atoms are carbon, particularly other NHC ligands, may be a powerful strategy for improving both the efficiency and device lifetimes of deep blue PHOLEDs.

## SUPPLEMENTARY MATERIAL

Listings of the Cartesian coordinates of the optimized ground-state geometry of the four NHC Ir(III) complexes studied in this work.

## ACKNOWLEDGMENTS

We thank Thilini Batagoda from the group of Mark E. Thompson for helpful communications. X. Z. is supported by a University of Queensland Development Research Fellowship (2017-2020). B. J. P. was supported by an ARC Future Fellowship (FT130100161). This research was undertaken with the assistance of resources provided at the NCI National Facility systems at the Australian National University through the National Computational Merit Allocation Scheme supported by the Australian Government, including grant LE120100181.